# NUMERICAL PERFORMANCE OF COMPACT FOURTH ORDER FORMULATION OF THE NAVIER-STOKES EQUATIONS

**Ercan Erturk**

Gebze Institute of Technology, Energy Systems Engineering Department,
Gebze, Kocaeli 41400, Turkey

In this study the numerical performance of the fourth order compact formulation of the steady 2-D incompressible Navier-Stokes equations introduced by Erturk *et al*. (*Int. J. Numer. Methods Fluids*, **50**, 421-436) will be presented. The benchmark driven cavity flow problem will be solved using the introduced compact fourth order formulation of the Navier-Stokes equations with two different line iterative semi-implicit methods for both second and fourth order spatial accuracy. The extra CPU work needed for increasing the spatial accuracy from second order ($\mathcal{O}(\Delta x^2)$) to fourth order ($\mathcal{O}(\Delta x^4)$) formulation will be presented.

## 1. INTRODUCTION

In Computational Fluid Dynamics (CFD) field of study High Order Compact formulations are becoming more popular. Compact formulations provide more accurate solutions in a compact stencil.

In finite difference, in order to achieve fourth order spatial accuracy, standard five point discretization can be used. When a five point discretization is used, the points near the boundaries have to be treated specially. Another way to achieve fourth order spatial accuracy is to use High Order Compact schemes. High Order Compact schemes provide fourth order spatial accuracy in a 3×3 stencil, hence this type of formulations can be used near the boundaries without a complexity.

In the literature, Zhang [14], Dennis and Hudson [3], MacKinnon and Johnson [6], Gupta *et al*. [5], Spotz and Carey [7] and Li *et al*. [9] have demonstrated the efficiency of high order compact schemes on the streamfunction and vorticity formulation of 2-D steady incompressible Navier-Stokes equations for uniform grids. Also in the literature the studies of Ge and Zhang [4] and Spotz and Carey [8] are example studies on the application of the high order compact scheme to nonuniform grids. The advantage of the high order compact schemes is that for a given flow problem and for a chosen grid mesh, high order compact formulation provides more accurate solutions ($\mathcal{O}(\Delta x^4)$) compared to standard second order formulation ($\mathcal{O}(\Delta x^2)$). Also for a given flow problem, the same level of accuracy of the solution obtained by standard second order formulation ($\mathcal{O}(\Delta x^2)$) using a certain grid mesh, can be obtained with a smaller grid mesh when high order compact ($\mathcal{O}(\Delta x^4)$) formulation is used.

Recently Erturk and Gokcol [11] have presented a new fourth order compact formulation. The uniqueness of this formulation is that the presented "Compact Fourth Order Formulation of the Navier-Stokes" (FONS) equations are in the same form with the Navier-Stokes (NS) equations with additional coefficients. In fact the

---

http://www.cavityflow.com

NS equations are a subset of the FONS equations and obtained when the additional coefficients are chosen as zero. Therefore any numerical method that solves the NS equations can be easily applied to the introduced FONS equations to obtain fourth order spatial accurate solutions. Moreover, the most important feature of the FONS equations is that any existing code with second order accuracy ($\mathcal{O}(\Delta x^2)$) can easily be changed to provide fourth order accuracy ($\mathcal{O}(\Delta x^4)$) by just adding some coefficients into the existing code at the expense of extra CPU work of evaluating these coefficients. This is an important feature such that if one already has a second order accurate code and wants to increase the accuracy of it to fourth order, instead of writing a new code, one can use the FONS equations and just by inserting some coefficient into the existing second order code, the existing second order code can turn into a fourth order code. Therefore the FONS equations introduced by Erturk and Gokcol [11] provide a very easy way to convert an existing second order accurate code into a fourth order accurate code and this is as simple as inserting some coefficients into the existing code. Of course, when this is done, the code will run a little slower because of the extra CPU work of evaluating the inserted coefficients. It will be good to estimate the CPU time needed for convergence of a converted fourth order accurate code compared to the CPU time needed for a second order accurate code.

Zhang [15] have studied the convergence and performance of iterative methods with fourth order compact discretization schemes. To the best of the author's knowledge in the literature there is not a study that documents the numerical performance of high order compact formulation the Navier-Stokes equations compared to regular second order formulation of the Navier-Stokes equations in terms of numerical stability and convergence for a chosen iterative method. In this study using the FONS equations introduced by Erturk and Gokcol [11], we will numerically solve the Navier-Stokes equations for both fourth order ($\mathcal{O}(\Delta x^4)$) and second order ($\mathcal{O}(\Delta x^2)$) spatial accuracy. This way we will be able to compare the convergence and stability characteristics of both formulations. In this study we will also document the extra CPU work that is needed for convergence when a second order accurate code is converted into a fourth order accurate code using the introduced FONS equations by Erturk and Gokcol [11]. The stability and convergence characteristics of both formulations and also the extra CPU work can show variation depending on the iterative numerical method used for the solution therefore in this study we will use two different line iterative semi-implicit numerical methods. Using these two numerical methods we will solve the benchmark driven cavity flow problem. First we will solve the cavity flow with second order ($\mathcal{O}(\Delta x^2)$) spatial accuracy then we will solve the same flow with fourth order ($\mathcal{O}(\Delta x^4)$) spatial accuracy. We will document the stability characteristics, such as the maximum allowable time increment ($\Delta t$), convergence characteristics, such as the number of iterations and the CPU time necessary for a chosen convergence criteria and also the extra CPU work that is needed to increase the spatial accuracy of the numerical solution from second order to fourth order using the FONS equations.

## 2. FOURTH ORDER COMPACT FORMULATION

In non-dimensional form, steady 2-D incompressible Navier-Stokes equations in

---

http://www.cavityflow.com

streamfunction ($\psi$) and vorticity ($\omega$) formulation are given as

$$\frac{\partial^2 \psi}{\partial x^2} + \frac{\partial^2 \psi}{\partial y^2} = -\omega \qquad (1)$$

$$\frac{1}{Re}\frac{\partial^2 \omega}{\partial x^2} + \frac{1}{Re}\frac{\partial^2 \omega}{\partial y^2} = \frac{\partial \psi}{\partial y}\frac{\partial \omega}{\partial x} - \frac{\partial \psi}{\partial x}\frac{\partial \omega}{\partial y} \qquad (2)$$

where $x$ and $y$ are the Cartesian coordinates and $Re$ is the Reynolds number.

Erturk and Gokcol [11] have introduced the Fourth Order Navier-Stokes (FONS) equations. The introduced FONS equations are the following

$$\frac{\partial^2 \psi}{\partial x^2} + \frac{\partial^2 \psi}{\partial y^2} = -\omega + A \qquad (3)$$

$$\frac{1}{Re}(1+B)\frac{\partial^2 \omega}{\partial x^2} + \frac{1}{Re}(1+C)\frac{\partial^2 \omega}{\partial y^2} = \left(\frac{\partial \psi}{\partial y} + D\right)\frac{\partial \omega}{\partial x} - \left(\frac{\partial \psi}{\partial x} + E\right)\frac{\partial \omega}{\partial y} + F \qquad (4)$$

where

$$A = -\frac{\Delta x^2}{12}\frac{\partial^2 \omega}{\partial x^2} - \frac{\Delta y^2}{12}\frac{\partial^2 \omega}{\partial y^2} - \left(\frac{\Delta x^2}{12} + \frac{\Delta y^2}{12}\right)\frac{\partial^4 \psi}{\partial x^2 \partial y^2}$$

$$B = -Re\frac{\Delta x^2}{6}\frac{\partial^2 \psi}{\partial x \partial y} + Re^2\frac{\Delta x^2}{12}\frac{\partial \psi}{\partial y}\frac{\partial \psi}{\partial y}$$

$$C = Re\frac{\Delta y^2}{6}\frac{\partial^2 \psi}{\partial x \partial y} + Re^2\frac{\Delta y^2}{12}\frac{\partial \psi}{\partial x}\frac{\partial \psi}{\partial x}$$

$$D = \left(\frac{\Delta x^2}{12} + \frac{\Delta y^2}{12}\right)\frac{\partial^3 \psi}{\partial x^2 \partial y} - Re\frac{\Delta x^2}{12}\frac{\partial \psi}{\partial y}\frac{\partial^2 \psi}{\partial x \partial y} + Re\frac{\Delta y^2}{12}\frac{\partial \psi}{\partial x}\frac{\partial^2 \psi}{\partial y^2}$$

$$E = \left(\frac{\Delta x^2}{12} + \frac{\Delta y^2}{12}\right)\frac{\partial^3 \psi}{\partial x \partial y^2} - Re\frac{\Delta x^2}{12}\frac{\partial \psi}{\partial y}\frac{\partial^2 \psi}{\partial x^2} + Re\frac{\Delta y^2}{12}\frac{\partial \psi}{\partial x}\frac{\partial^2 \psi}{\partial x \partial y}$$

$$F = \left(\frac{\Delta x^2}{12} + \frac{\Delta y^2}{12}\right)\frac{\partial \psi}{\partial y}\frac{\partial^3 \omega}{\partial x \partial y^2} - \left(\frac{\Delta x^2}{12} + \frac{\Delta y^2}{12}\right)\frac{\partial \psi}{\partial x}\frac{\partial^3 \omega}{\partial x^2 \partial y} - \frac{\Delta x^2}{6}\frac{\partial^2 \psi}{\partial x^2}\frac{\partial^2 \omega}{\partial x \partial y}$$

$$+ \frac{\Delta y^2}{6}\frac{\partial^2 \psi}{\partial y^2}\frac{\partial^2 \omega}{\partial x \partial y} + Re\left(\frac{\Delta x^2}{12} + \frac{\Delta y^2}{12}\right)\frac{\partial \psi}{\partial x}\frac{\partial \psi}{\partial y}\frac{\partial^2 \omega}{\partial x \partial y} - \left(\frac{\Delta x^2}{12} - \frac{\Delta y^2}{12}\right)\frac{\partial \omega}{\partial x}\frac{\partial \omega}{\partial y}$$

$$- \frac{1}{Re}\left(\frac{\Delta x^2}{12} + \frac{\Delta y^2}{12}\right)\frac{\partial^4 \omega}{\partial x^2 \partial y^2} \qquad (5)$$

As it is described briefly in Erturk and Gokcol [11], the numerical solutions of FONS equations (3) and (4) are fourth order accurate to NS equations (1) and (2), *strictly* provided that second order central discretizations shown in Table 1 are used and also *strictly* provided that a uniform grid mesh with $\Delta x$ and $\Delta y$ is used. We note that NS equations are a subset of FONS equations and obtained when the coefficients $A$, $B$, $C$, $D$, $E$ and $F$ in FONS equations are chosen as zero. The FONS equations are in the same form with the NS equations therefore any iterative numerical method applied to streamfunction and vorticity equation (1) and (2) can be easily applied to fourth order streamfunction and vorticity equation (3) and (4). Moreover if there is an existing code that solves the streamfunction and vorticity equation (1) and (2) with second order spatial accuracy, by just adding some coefficients $A$, $B$, $C$, $D$, $E$ and $F$ into the existing code using the FONS equations, the same existing code can

---

provide fourth order spatial accuracy. A single numerical code for the solution of the FONS equations can provide both second order and fourth order spatial accuracy by just setting some coefficients. Of course there is a pay off switching from second order to fourth order spatial accuracy, that is the extra cost of CPU work of calculating the coefficients $A$, $B$, $C$, $D$, $E$ and $F$ as defined in equation (5).

## 3. FINITE DIFFERENCE EQUATIONS

For numerical solutions of the Navier-Stokes equations (1) and (2), the following finite difference equations provide second order ($\mathcal{O}(\Delta x^2)$) accuracy

$$\psi_{xx} + \psi_{yy} = -\omega \tag{6}$$

$$\frac{1}{Re}\omega_{xx} + \frac{1}{Re}\omega_{yy} = \psi_y \omega_x - \psi_x \omega_y \tag{7}$$

where subscripts denote derivatives as defined in Table 1.

As explained in Erturk and Gokcol [11] the solution of the following finite difference equations are fourth order ($\mathcal{O}(\Delta x^4)$) accurate to the Navier-Stokes.

$$\psi_{xx} + \psi_{yy} = -\omega + A \tag{8}$$

$$\frac{1}{Re}(1+B)\omega_{xx} + \frac{1}{Re}(1+C)\omega_{yy} = (\psi_y + D)\omega_x - (\psi_x + E)\omega_y + F \tag{9}$$

where

$$A = -\frac{\Delta x^2}{12}\omega_{xx} - \frac{\Delta y^2}{12}\omega_{yy} - \left(\frac{\Delta x^2}{12} + \frac{\Delta y^2}{12}\right)\psi_{xxyy}$$

$$B = -Re\frac{\Delta x^2}{6}\psi_{xy} + Re^2\frac{\Delta x^2}{12}\psi_y\psi_y$$

$$C = Re\frac{\Delta y^2}{6}\psi_{xy} + Re^2\frac{\Delta y^2}{12}\psi_x\psi_x$$

$$D = \left(\frac{\Delta x^2}{12} + \frac{\Delta y^2}{12}\right)\psi_{xxy} - Re\frac{\Delta x^2}{12}\psi_y\psi_{xy} + Re\frac{\Delta y^2}{12}\psi_x\psi_{yy}$$

$$E = \left(\frac{\Delta x^2}{12} + \frac{\Delta y^2}{12}\right)\psi_{xyy} - Re\frac{\Delta x^2}{12}\psi_y\psi_{xx} + Re\frac{\Delta y^2}{12}\psi_x\psi_{xy}$$

$$F = \left(\frac{\Delta x^2}{12} + \frac{\Delta y^2}{12}\right)\psi_y\omega_{xyy} - \left(\frac{\Delta x^2}{12} + \frac{\Delta y^2}{12}\right)\psi_x\omega_{xxy} - \frac{\Delta x^2}{6}\psi_{xx}\omega_{xy}$$

$$+ \frac{\Delta y^2}{6}\psi_{yy}\omega_{xy} + Re\left(\frac{\Delta x^2}{12} + \frac{\Delta y^2}{12}\right)\psi_x\psi_y\omega_{xy} - \left(\frac{\Delta x^2}{12} - \frac{\Delta y^2}{12}\right)\omega_x\omega_y$$

$$- \frac{1}{Re}\left(\frac{\Delta x^2}{12} + \frac{\Delta y^2}{12}\right)\omega_{xxyy} \tag{10}$$

Note that equations (8) and (9) are in the same form with equations (6) and (7) except with additional coefficients $A$, $B$, $C$, $D$, $E$ and $F$. In equations (8) and (9) if the coefficients $A$, $B$, $C$, $D$, $E$ and $F$ are chosen to be zero then the solution of these equations are second order accurate ($\mathcal{O}(\Delta x^2, \Delta y^2)$) to NS equations, since when the coefficients are zero, equations (8) and (9) are identical with equations (6) and (7).

However, in equations (8) and (9) if the coefficients $A$, $B$, $C$, $D$, $E$ and $F$ are calculated as they are defined in equation (10), then the solution of these equations (8) and (9) are fourth order accurate ($\mathcal{O}(\Delta x^4, \Delta x^2 \Delta y^2, \Delta y^4)$) to NS equations. When FONS equations are used, one can easily switch from second order to fourth order spatial accuracy using a single equation by just using the appropriate coefficients. Computationally, calculating the coefficients defined in equation (10) when fourth order accuracy is desired will require an extra CPU work compared to second order accuracy. In order to quantify the extra CPU work to switch from equations (6) and (7) to equations (8) and (9), i.e. switch from second order accuracy to fourth order accuracy, we solve the above equations with two different line iterative semi-implicit numerical method and document the CPU time for comparison.

We note that the equations (8) and (9) are nonlinear equations therefore they need to be solved in an iterative manner. In order to have an iterative numerical algorithm we assign pseudo time derivatives to equations (8) and (9), thus we have

$$\frac{\partial \psi}{\partial t} = \psi_{xx} + \psi_{yy} + \omega - A \tag{11}$$

$$\frac{\partial \omega}{\partial t} = \frac{1}{Re}(1+B)\omega_{xx} + \frac{1}{Re}(1+C)\omega_{yy} - (\psi_y + D)\omega_x + (\psi_x + E)\omega_y - F \tag{12}$$

We solve these equations (11) and (12) in the pseudo time domain until the solution converges to steady state.

One of the numerical methods we will use to solve equations (11) and (12) is the Alternating Direction Implicit (ADI) method. ADI method is a very widely used numerical method and in this method a two dimensional problem is solved in two sweeps while solving the equation implicitly in one dimension in each sweep. The reader is referred to [12], [1] and [2] for details. When we apply the ADI method to solve equation (11), first we solve the following tri-diagonal system in $x$-direction

$$\left(1 - \frac{\Delta t}{2}\delta_{xx}\right)\psi^{n+\frac{1}{2}} = \psi^n + \frac{\Delta t}{2}\psi_{yy}^n + \frac{\Delta t}{2}\omega - \frac{\Delta t}{2}A \tag{13}$$

then we solve the following tri-diagonal system in $y$-direction

$$\left(1 - \frac{\Delta t}{2}\delta_{yy}\right)\psi^{n+1} = \psi^{n+\frac{1}{2}} + \frac{\Delta t}{2}\psi_{xx}^{n+\frac{1}{2}} + \frac{\Delta t}{2}\omega - \frac{\Delta t}{2}A \tag{14}$$

Similarly when we apply the ADI method to solve equation (12), we first solve the following tri-diagonal system in $x$-direction

$$\left(1 - \frac{\Delta t}{2}\frac{1}{Re}(1+B)\delta_{xx} + \frac{\Delta t}{2}(\psi_y + D)\delta_x\right)\omega^{n+\frac{1}{2}} = \omega^n$$
$$+ \frac{\Delta t}{2}(1+C)\frac{1}{Re}\omega_{yy}^n + \frac{\Delta t}{2}(\psi_x + E)\omega_y^n - \frac{\Delta t}{2}F \tag{15}$$

then we solve the following tri-diagonal system in $y$-direction

$$\left(1 - \frac{\Delta t}{2}\frac{1}{Re}(1+C)\delta_{yy} - \frac{\Delta t}{2}(\psi_x + E)\delta_y\right)\omega^{n+1} = \omega^{n+\frac{1}{2}}$$
$$+ \frac{\Delta t}{2}\frac{1}{Re}(1+B)\omega_{xx}^{n+\frac{1}{2}} - \frac{\Delta t}{2}(\psi_y + D)\omega_x^{n+\frac{1}{2}} - \frac{\Delta t}{2}F \tag{16}$$

where $\delta_{xx}$ and $\delta_{yy}$ denote the second order finite difference operators, and similarly $\delta_x$ and $\delta_y$ denote the first order finite difference operators in $x$- and $y$-direction respectively, for example

$$\delta_x \theta = \frac{\theta_{i+1,j} - \theta_{i-1,j}}{2\Delta x} + \mathcal{O}(\Delta x^2)$$

$$\delta_{xx} \theta = \frac{\theta_{i+1,j} - 2\theta_{i,j} + \theta_{i-1,j}}{\Delta x^2} + \mathcal{O}(\Delta x^2) \qquad (17)$$

where $i$ and $j$ are the grid index and $\theta$ denote any differentiable quantity.

The second numerical method we will use is the efficient numerical method proposed by Erturk *et al.* [10]. Following Erturk *et al.* [10], first using an implicit Euler approximation for the pseudo time derivatives in equations (11) and (12), we obtain the following finite difference formulations

$$\left(1 - \Delta t\, \delta_{xx} - \Delta t \delta_{yy}\right)\psi^{n+1} = \psi^n + \Delta t \omega^n - \Delta t A^n \qquad (18)$$

$$\left(1 - \Delta t \frac{1}{Re}(1+B)\delta_{xx} - \Delta t \frac{1}{Re}(1+C)\delta_{yy} \right.$$
$$\left. + \Delta t(\psi_y + D)\delta_x - \Delta t(\psi_x + E)\delta_y \right)\omega^{n+1} = \omega^n - \Delta t F^n \qquad (19)$$

We note that equations (18) and (19) are in fully implicit form and each equation requires the solution of a large banded matrix which is not computationally efficient. Instead, we spatially factorize these equations (18) and (19) thus we obtain the following finite difference equations

$$\left(1 - \Delta t\, \delta_{xx}\right)\left(1 - \Delta t\, \delta_{yy}\right)\psi^{n+1} = \psi^n + \Delta t \omega^n - \Delta t A^n \qquad (20)$$

$$\left(1 - \Delta t \frac{1}{Re}(1+B)\delta_{xx} + \Delta t(\psi_y + D)\delta_x\right) \times$$
$$\left(1 - \Delta t \frac{1}{Re}(1+C)\delta_{yy} - \Delta t(\psi_x + E)\delta_y\right)\omega^{n+1} = \omega^n - \Delta t F^n \qquad (21)$$

The advantage of these equations (20) and (21) is that each equation requires the solution of tridiagonal systems which is computationally very efficient using the Thomas algorithm. However, spatial factorization introduces $\Delta t^2$ terms into the left hand side of equations (20) and (21), also these terms remain in the solution even at the steady state. To cancel out these $\Delta t^2$ terms due to the factorization, Erturk *et al.* [10] have added the same amount of $\Delta t^2$ terms to the right hand side of the equations so that the equations recover the correct physical representation at the steady state. The final form of the finite difference equations take the following form

$$\left(1 - \Delta t\, \delta_{xx}\right)\left(1 - \Delta t\, \delta_{yy}\right)\psi^{n+1} = \psi^n + \Delta t \omega^n - \Delta t A^n + \left(\Delta t\, \delta_{xx}\right)\left(\Delta t\, \delta_{yy}\right)\psi^n \qquad (22)$$

$$\left(1 - \Delta t \frac{1}{Re}(1+B)\delta_{xx} + \Delta t(\psi_y + D)\delta_x\right) \times$$
$$\left(1 - \Delta t \frac{1}{Re}(1+C)\delta_{yy} - \Delta t(\psi_x + E)\delta_y\right)\omega^{n+1} = \omega^n - \Delta t F^n$$
$$\left(\Delta t \frac{1}{Re}(1+B)\delta_{xx} - \Delta t(\psi_y + D)\delta_x\right) \times$$

$$\left(\Delta t \frac{1}{Re}(1+C)\delta_{yy} + \Delta t(\psi_x + E)\delta_y\right)\omega^n \quad (23)$$

The reader is referred to Erturk *et al*. [10] for details. The solution methodology for equations (22) and (23) involves a two stage time level updating. For example, for the solution of equation (22), we first solve for the introduced variable $f$ in $x$-direction in the following tri-diagonal system

$$\left(1 - \Delta t\, \delta_{xx}\right)f = \psi^n + \Delta t\omega^n - \Delta t A^n + \left(\Delta t\, \delta_{xx}\right)\left(\Delta t\, \delta_{yy}\right)\psi^n \quad (24)$$

When the solution for $f$ is obtained, the streamfunction variable is advanced into the next time level by solving the following tri-diagonal system in $y$-direction

$$\left(1 - \Delta t\, \delta_{yy}\right)\psi^{n+1} = f \quad (25)$$

Similarly, for the solution of equation (23), we first solve for the introduced variable $g$ in $x$-direction in the following tri-diagonal system

$$\left(1 - \Delta t \frac{1}{Re}(1+B)\delta_{xx} + \Delta t(\psi_y + D)\delta_x\right)g = \omega^n - \Delta t F^n$$

$$\left(\Delta t \frac{1}{Re}(1+B)\delta_{xx} - \Delta t(\psi_y + D)\delta_x\right) \times$$

$$\left(\Delta t \frac{1}{Re}(1+C)\delta_{yy} + \Delta t(\psi_x + E)\delta_y\right)\omega^n \quad (26)$$

When the solution for $g$ is obtained, the vorticity variable is advanced into the next time level by solving the following tri-diagonal system in $y$-direction

$$\left(1 - \Delta t \frac{1}{Re}(1+C)\delta_{yy} - \Delta t(\psi_x + E)\delta_y\right)\omega^{n+1} = g \quad (27)$$

Störtkuhl *et al*. [13] have presented an analytical asymptotic solution near the corners of cavity and using finite element bilinear shape functions they also have presented a singularity removed boundary condition for vorticity at the corner points as well as at the wall points. For the boundary conditions, in both of the numerical methods described above we follow Störtkuhl *et al*. [13] and use the following expression for calculating vorticity values at the wall

$$\frac{1}{3\Delta h^2}\begin{bmatrix} \bullet & \bullet & \bullet \\ \frac{1}{2} & -4 & \frac{1}{2} \\ 1 & 1 & 1 \end{bmatrix}\psi + \frac{1}{9}\begin{bmatrix} \bullet & \bullet & \bullet \\ \frac{1}{2} & 2 & \frac{1}{2} \\ \frac{1}{4} & 1 & \frac{1}{4} \end{bmatrix}\omega = -\frac{V}{\Delta h} \quad (28)$$

where $V$ is the speed of the wall which is equal to 1 for the moving top wall and equal to 0 for the three stationary walls.

For corner points, we again follow Störtkuhl *et al*. [13] and use the following expression for calculating the vorticity values

$$\frac{1}{3\Delta h^2}\begin{bmatrix} \bullet & \bullet & \bullet \\ \bullet & -2 & \frac{1}{2} \\ \bullet & \frac{1}{2} & 1 \end{bmatrix}\psi + \frac{1}{9}\begin{bmatrix} \bullet & \bullet & \bullet \\ \bullet & 1 & \frac{1}{2} \\ \bullet & \frac{1}{2} & \frac{1}{4} \end{bmatrix}\omega = -\frac{V}{2\Delta h} \qquad (29)$$

where again $V$ is equal to 1 for the upper two corners and it is equal to 0 for the bottom two corners.

In explicit notation, for the wall points shown in Figure 1, the vorticity is calculated as the following

$$\omega_b = -\frac{9V}{2\Delta h} - \frac{3}{2\Delta h^2}(\psi_d + \psi_e + \psi_f) - \frac{1}{8}(2\omega_a + 2\omega_c + \omega_d + 4\omega_e + \omega_f) \qquad (30)$$

Similarly, for the corner points also shown in Figure 1, the vorticity is calculated as the following.

$$\omega_b = -\frac{9V}{2\Delta h} - \frac{3}{\Delta h^2}\psi_f - \frac{1}{4}(2\omega_c + \omega_f + 2\omega_e) \qquad (31)$$

The reader is referred to Störtkuhl *et al*. [13] for details on the boundary conditions

## 4. RESULTS

In order to quantify the extra CPU work needed when a second order accuracy code is converted into a fourth order accuracy code using the FONS equations introduced by Erturk and Gokcol [11], we have solved both the second order and fourth order accurate equations (6), (7), (8) and (9) for the solution of the driven cavity flow. We consider the driven cavity flow for Reynolds numbers of $Re$ =100, 1000 and 3200, with using a grid mesh of $128 \times 128$ ($\Delta x = \Delta y = \Delta h$). We note that since we are mainly interested in finding the ratio of CPU time needed for convergence of a fourth order accuracy code to CPU time needed for convergence of a second order accuracy code, the choice of grid mesh size is not important, such that the ratio will be the same whether a coarse or fine grid mesh is used.

In solving the equations we decided to use the Alternating Direction Implicit (ADI) method and the numerical method proposed by Erturk *et al*. [10]. By using two different numerical methods we would be able to see if the extra CPU work is dependent on the numerical method used. While doing this, as a by-product, we would also be able to compare the ADI method and the method proposed by Erturk *et al*. [10] in terms of numerical performance. In both of the numerical method we use, for both second order and fourth order accuracy, the two equations, i.e. the streamfunction and the vorticity equations, are solved separately. In order to document the extra CPU work when a fourth order accuracy is desired what we do is, we first solve for second order accuracy and solve equations (6) and (7). Then keeping the number of grids, the time step $\Delta t$ and boundary conditions the same, we solve for fourth order accuracy thus we calculate and insert the coefficients $A$, $B$, $C$, $D$, $E$ and $F$ into the equations and solve for equations (8) and (9). While we solve the same flow problem, i.e. the driven cavity flow, for both second and fourth order accuracy we document the necessary number of iterations and the CPU time needed for a certain defined convergence criteria. This way we would be able to compare the

---

http://www.cavityflow.com

convergence characteristics of both second and fourth order formulations in terms of the number of iterations and the CPU time, and also we would be able to document the extra CPU time needed if a second order code is converted into a fourth order code using the FONS equations.

For the choice of the time steps in solving the governing equations, we decided to use different time steps, $\Delta t$, for streamfunction and vorticity equations. In both of the numerical methods we use, while solving both the streamfunction and vorticity equations, tri-diagonal matrices appear on the implicit LHS of the equations. When second order accuracy is considered, in streamfunction equation the diagonal elements on the LHS matrices become $1 + \frac{2\Delta t}{\Delta h^2}$, also in vorticity equations the diagonal elements on the LHS matrices become $1 + \frac{1}{Re}\frac{2\Delta t}{\Delta h^2}$. We choose different time steps for streamfunction and vorticity equations that would make the diagonal elements the same in both equations. Therefore for streamfunction equation we use $\Delta t = \alpha \Delta h^2$ and for vorticity equation we use $\Delta t = \alpha Re \Delta h^2$, where $\alpha$ is a coefficient we can choose. For fourth order accuracy we use the same time steps we use in second order accuracy. By using the same time steps in second and fourth order accuracy we would be able to compare the numerical stability of the FONS equations and the NS equations. In order to do this first we both solve the NS and the FONS equations using the same time steps. Then we increase $\alpha$, i.e. increase the time step $\Delta t$, and solve the NS and the FONS equations again. We continue doing this until at some $\Delta t$ the solution does not converge. Therefore we would document the maximum allowable $\Delta t$ for convergence for both the NS and the FONS equations for a given Reynolds number and grid mesh. This maximum allowable $\Delta t$ for convergence is an indicative of the numerical stability. For example, using either of the numerical method, i.e. the ADI method or the Erturk method, we solve the same flow problem using both second and fourth order formulations. Therefore for a chosen numerical method, the maximum allowable time step for second and fourth order formulations will be indicative of the numerical stability characteristics of the second order formulation compared to that of the fourth order formulation. Also, using either of the formulations, i.e. second or fourth order formulations, we solve the same flow problem using both the ADI method and the Erturk method. Therefore, for a chosen formulation, the maximum allowable time step for the ADI and the Erturk methods will be indicative of the numerical stability characteristics of the ADI method compared to that of the Erturk method.

Our extensive numerical studies show that the increase in the extra CPU work is dependent on the computer and the compiler used. In this study, we run the codes on a 64 bit HP ES45 machine with EV68 AlphaChip 1.25 GHz processors with HP Tru64 UNIX operating system. We run the codes with both compiled normally and also compiled using maximum compiler optimization (`-fast -O5`).

We start the iterations from a homogenous initial guess and continue until a certain condition of convergence is satisfied. As the measure of the convergence to the same level, the residual of the equations can be used as it was also used in Erturk *et al*. [10]. However we are solving two different equations, the NS and the FONS equations, and try to compare the CPU time of convergence for each equation to the same level. Therefore the residual of these equations may not show the same convergence level. Alternatively, we can use the difference of the streamfunction and vorticity variables

---

between two time steps as the measure of convergence. However the solutions of the two different equations are slightly different since one is spatially second order and the other is fourth order accurate. Since the solutions are different, the difference of the streamfunction and vorticity variables between two time steps may not also show the same convergence level for those equations. Therefore as the convergence criteria we decided to use the difference of the streamfunction and vorticity variables between two time steps normalized by the previous value of the corresponding variable, such that

$$\text{Residual}_\psi = \max\left(\left|\frac{\psi_{i,j}^{n+1} - \psi_{i,j}^n}{\psi_{i,j}^n}\right|\right)$$

$$\text{Residual}_\omega = \max\left(\left|\frac{\omega_{i,j}^{n+1} - \omega_{i,j}^n}{\omega_{i,j}^n}\right|\right) \quad (32)$$

These residuals provide an indication of the maximum percent change in $\psi$ and $\omega$ variables in each iteration step. In all of the data presented in this study, for both the solutions of the NS and the FONS equations, obtained using both of the numerical methods, we let the iterations converge until both $\text{Residual}_\psi$ and $\text{Residual}_\omega$ are less than $10^{-8}$. At this convergence level, this would indicate that the variables $\psi$ and $\omega$ are changing less than 0.000001% of their value between two iterations at every grid point in the mesh.

Figure 2, 3 and 4 show the streamline and vorticity contours of the driven cavity flow for $Re$ =100, 1000 and 3200 respectively, obtained using the method proposed by Erturk *et al.* [10] applied to FONS equations ($\mathcal{O}(\Delta x^4)$). We note that both second order and fourth order accurate solutions of ADI method and the Erturk method, agree well with the solutions found in the literature especially with Erturk *et al.* [10],[11].

Using both of the numerical methods, we solve the driven cavity flow using different coefficients for time ($\alpha$), i.e. using different time steps, and document the CPU time and iteration number needed to the desired convergence level explained above. Table 2 shows the CPU time and iteration numbers for ADI method for different Reynolds numbers using various $\alpha$ values. Table 3 shows the same for the method proposed by Erturk *et al.* [10]. Looking at Table 2 and 3, for both of the numerical methods the number of iterations for convergence is almost the same for second order and fourth order accuracy. However for both of the numerical methods the CPU time for fourth order accuracy is greater than the CPU time for second order accuracy as expected since the coefficients $A$, $B$, $C$, $D$, $E$ and $F$ have to be calculated at each iteration in fourth order accuracy which will result in an increase in the CPU time. The ratio of the CPU times of fourth order accuracy to second order accuracy show the increase in CPU time when we switch from second order accuracy to fourth order accuracy. From Tables 2 and 3 we see that this ratio seems to increase slightly when Reynolds number increases.

It seems that for both of the numerical methods, at a given Reynolds number, the ratios of CPU time and iteration number for second and fourth order accuracy is constant and it is independent of the time step, i.e. $\alpha$.

For the ADI method, in Table 2, when the order of accuracy is increased to fourth order from second order, the CPU time increases almost 1.76 times for $Re$=100. This number increases as the Reynolds number increases and the increase in CPU time becomes 1.90 times for $Re$=3200.

For the method proposed by Erturk et al. [10], in Table 3, when the order of accuracy is increased to fourth order from second order, the CPU time increases almost 1.60 times for $Re$=100 and it is almost 1.68 times for $Re$=3200.

We note that, when the order of accuracy is increased from second order to fourth order, the 1.6 and 1.68 times increase in CPU time for $Re$=100 and 3200 respectively for the method proposed by Erturk et al. [10] are less than the equivalent 1.76 and 1.90 times increase in CPU time for the same Reynolds numbers for the ADI method. This shows that, the extra CPU time needed for fourth order accuracy when FONS equations are used, is dependent on the numerical method used and the extra CPU time for the method proposed by Erturk et al. [10] is much lower than the extra CPU time for the ADI method.

In Table 2 and 3, comparing the two methods, for the *same* Reynolds numbers and for the *same* $\alpha$ values ($\Delta t$), the iteration numbers for convergence are almost the same for both ADI and the method proposed by Erturk et al. [10], however the CPU time for ADI is less than that of the method proposed by Erturk et al. [10]. The reason for this is that in the method proposed by Erturk et al. [10] on the RHS of the finite difference equations more terms have to be calculated at each iteration and this increases the CPU time compared to the ADI method.

For faster convergence one can use larger time steps, if the numerical method used has a higher numerical stability limit. Therefore, for a numerical method, the maximum allowable time step ($\Delta t$) for convergence gives an indication of the numerical stability limit of the numerical method. Since in this study we have used two different numerical methods for the same flow problem, we decided to compare the numerical stability limit of the two methods applied to both the NS and the FONS equations, by finding the maximum allowable time step for convergence for both numerical methods. In order to find the maximum allowable time step for convergence, for a given Reynolds number we solve the second and fourth order equations using both of the numerical methods several times while increasing $\alpha$ with 0.01 increments each time, until the solution no longer converges.

For the ADI method the maximum allowable $\alpha$ for convergence is 0.79 for second order accuracy and it is 0.78 for fourth order accuracy for $Re$=1000. For the method proposed by Erturk et al. [10] the maximum $\alpha$ values was 1.89 for second order accuracy and it was 1.75 for fourth order accuracy. This would indicate that one can use much larger time steps in Erturk method compared to the ADI method, for example, for $Re$=1000 the method proposed by Erturk et al. [10] allows to use 2.4 times larger time step for the NS equations and 2.2 times larger time step for the FONS equations than the ADI method. From this we can conclude that the Erturk method has better numerical stability characteristics compared to the ADI method. When the maximum allowable time steps are used, the required CPU time for the method proposed by Erturk et al. [10] is almost 0.53 of the required CPU time for the ADI method. This means that the method proposed by Erturk et al. [10] converge almost twice faster than the ADI method when the maximum allowable time steps are

---

used.

Comparing the numerical stability of the NS and the FONS equations, we see that for a chosen numerical method the FONS equations have slightly less stability limit than that of the NS equations. For example, for the ADI method and for $Re=1000$ the value of 0.79 for maximum allowable $\alpha$ for convergence for second order accuracy drop down to 0.78 when fourth order accuracy is used. Also, for the Erturk method and for $Re=1000$ the maximum allowable $\alpha$ value of 1.89 for second order accuracy drop down to 1.75 if we switch to fourth order accuracy. This would indicate that, for fourth order formulations the maximum allowable time step for convergence is lower than the maximum allowable time step for convergence for second order formulations.

We then decided to run the same codes compiled with using the maximum compiler optimization (`-fast -O5`). Table 4 and 5 document the CPU time and iteration numbers when compiler optimization is used for the same conditions documented in Table 2 and 3 respectively. Comparing the numbers in Table 4 and 2 for ADI method, compiler optimization decreases the necessary CPU time for convergence about 0.71 times for second order accuracy and about 0.51 times for fourth order accuracy. Also comparing the same in Table 5 and 3 for the method proposed by Erturk *et al*. [10], compiler optimization decreases the necessary CPU time for convergence about 0.54 and 0.45 for second order and fourth order accuracy respectively. These numbers show that compiler optimization decreases the CPU time significantly and for the numerical method proposed by Erturk *et al*. [10] the codes almost run twice faster in terms of CPU time when compiler optimization is used.

## 2. CONCLUSIONS

The FONS equations introduced by Erturk and Gokcol [11] are in the same form of the NS equations, therefore any numerical method that solve the Navier-Stokes equations can be easily applied to the FONS equations in order to obtain fourth order accurate solutions ($\mathcal{O}(\Delta x^4)$). One of the features of the introduced FONS equations is that any existing code that solve the Navier-Stokes equations with second order accuracy ($\mathcal{O}(\Delta x^2)$) can be easily altered to provide fourth order accuracy ($\mathcal{O}(\Delta x^4)$) just by adding some coefficients into the existing code using the FONS equations. This way, the accuracy of any second order code can be easily increased to fourth order, however there is a pay off for this increased accuracy, that is the extra CPU time for calculating the added coefficients.

In this study we have solved the NS equations and the FONS equations to document the extra CPU time necessary for convergence when an existing second order accurate code is altered to provide fourth order accurate solutions. For this we have used two different numerical methods. We find that the extra CPU time is slightly dependent on the numerical method used. For the ADI method to obtain fourth order accurate solutions of driven cavity flow, the CPU time increases 1.8 times compared to second order accurate solutions for $Re=1000$. Also for the numerical method proposed by Erturk *et al*. [10], with the cost of 1.64 times the CPU time necessary for second order accuracy, a fourth order accurate solutions can be obtained for $Re=1000$, using the FONS equations.

---

The FONS equations introduced by Erturk and Gokcol [11] provide a very easy way of obtaining fourth order accurate solutions by just adding some coefficients into an existing second order accurate code, at the expense of a minor increase in the CPU time.

**REFERENCES**
[1] Anderson DA, Tannehill JC, Pletcher RH. *Computational Fluid Mechanics and Heat Transfer*. McGraw-Hill, New York, 1984.
[2] Chung TJ. *Computational Fluid Dynamics*. Cambridge University Press, Cambridge, 2002.
[3] Dennis SC, Hudson JD. Compact $h^4$ Finite Difference Approximations to Operators of Navier-Stokes Type, *Journal of Computational Physics* 1989; **85**:390–416.
[4] Ge L, Zhang J. High Accuracy Iterative Solution of Convection Diffusion Equation with Boundary Layers on Nonuniform Grids, *Journal of Computational Physics* 2001; **171**:560–578.
[5] Gupta MM, Manohar RP, Stephenson JW. A Single Cell High Order Scheme for the Convection-Diffusion Equation with Variable Coefficients, *International Journal for Numerical Methods in Fluids* 1984; **4**:641–651.
[6] MacKinnon RJ, Johnson RW. Differential-Equation-Based Representation of Truncation Errors for Accurate Numerical Simulation, *International Journal for Numerical Methods in Fluids* 1991; **13**:739–757.
[7] Spotz WF, Carey GF. High-Order Compact Scheme for the Steady Streamfunction Vorticity Equations, *International Journal for Numerical Methods in Engineering* 1995; **38**:3497–3512.
[8] Spotz WF, Carey GF. Formulation and Experiments with High-Order Compact Schemes for Nonuniform Grids, *International Journal of Numerical Methods for Heat and Fluid Flow* 1998; **8**:288–303.
[9] Li M, Tang T, Fornberg B. A Compact Forth-Order Finite Difference Scheme for the Steady Incompressible Navier-Stokes Equations, *International Journal for Numerical Methods in Fluids* 1995; **20**:1137–1151.
[10] Erturk E, Corke TC, Gokcol C. Numerical Solutions of 2-D Steady Incompressible Driven Cavity Flow at High Reynolds Numbers, *International Journal for Numerical Methods in Fluids* 2005; **48**:747–774.
[11] Erturk E, Gokcol C. Fourth Order Compact Formulation of Navier-Stokes Equations and Driven Cavity Flow at High Reynolds Numbers, *International Journal for Numerical Methods in Fluids* 2006; **50**:421–436.
[12] Peaceman DW, Rachford Jr. HH. The numerical solution of parabolic and elliptic differential equations, *Journal of the Society for Industrial and Applied Mathematics* 1955; **3**:28–41.
[13] Störtkuhl T, Zenger C, Zimmer S. An Asymptotic Solution for the Singularity at the Angular Point of the Lid Driven Cavity, *International Journal of Numerical Methods for Heat and Fluid Flow* 1994; **4**:47–59.
[14] Zhang J. An Explicit Fourth-Order Compact Finite Difference Scheme for Three-Dimensional Convection-Diffusion Equation, *Communications in Numerical Methods in Engineering* 1998; **14**:209–218.
[15] Zhang J. On Convergence and Performance of Iterative Methods with Fourth-Order Compact Schemes, *Numerical Methods for Partial Differential Equations* 1998; **14**:263–280.

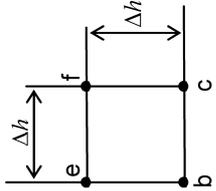
a) Wall points

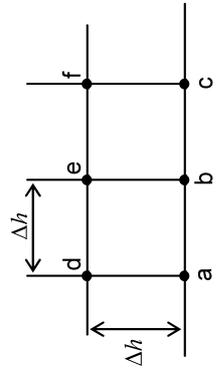
b) Corner points

Figure 1) Grid points at the wall and at the corner

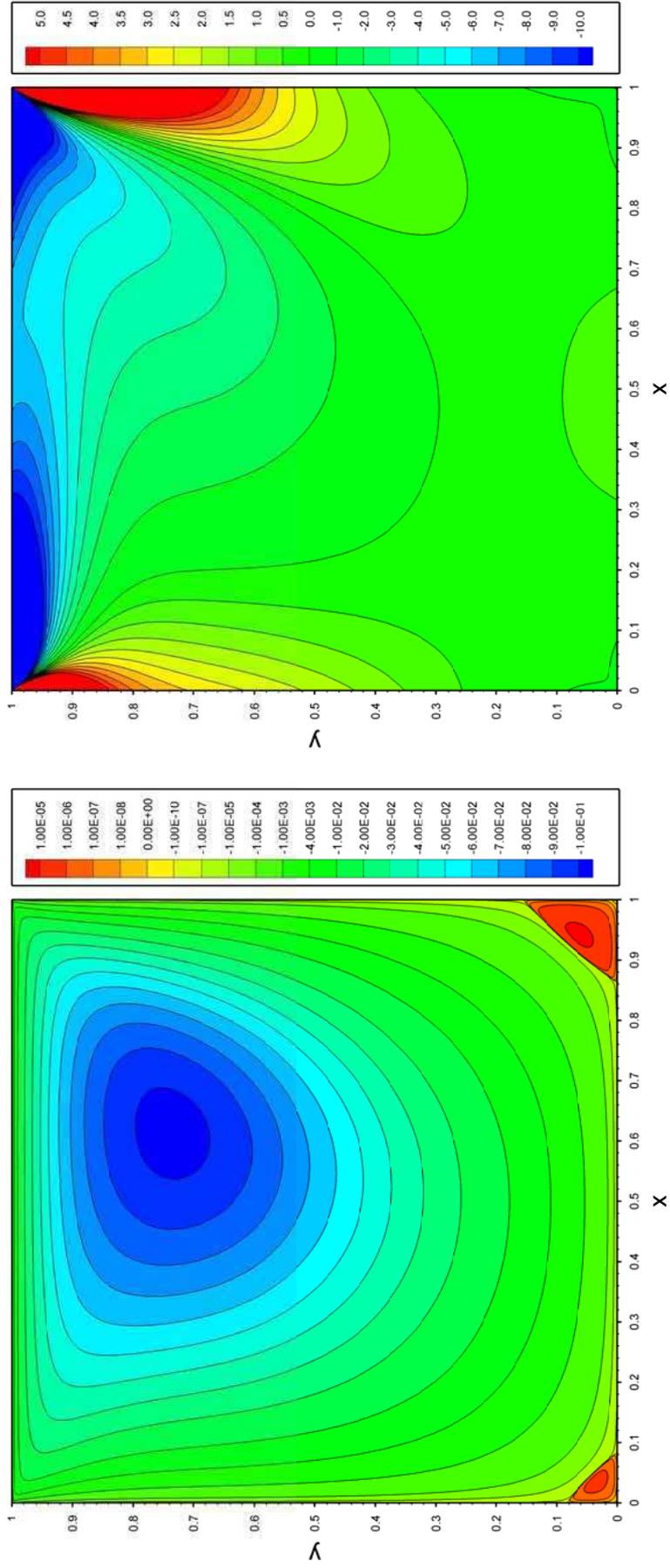

Figure 2) Streamline and vorticity contours of driven cavity flow, *Re*=100.

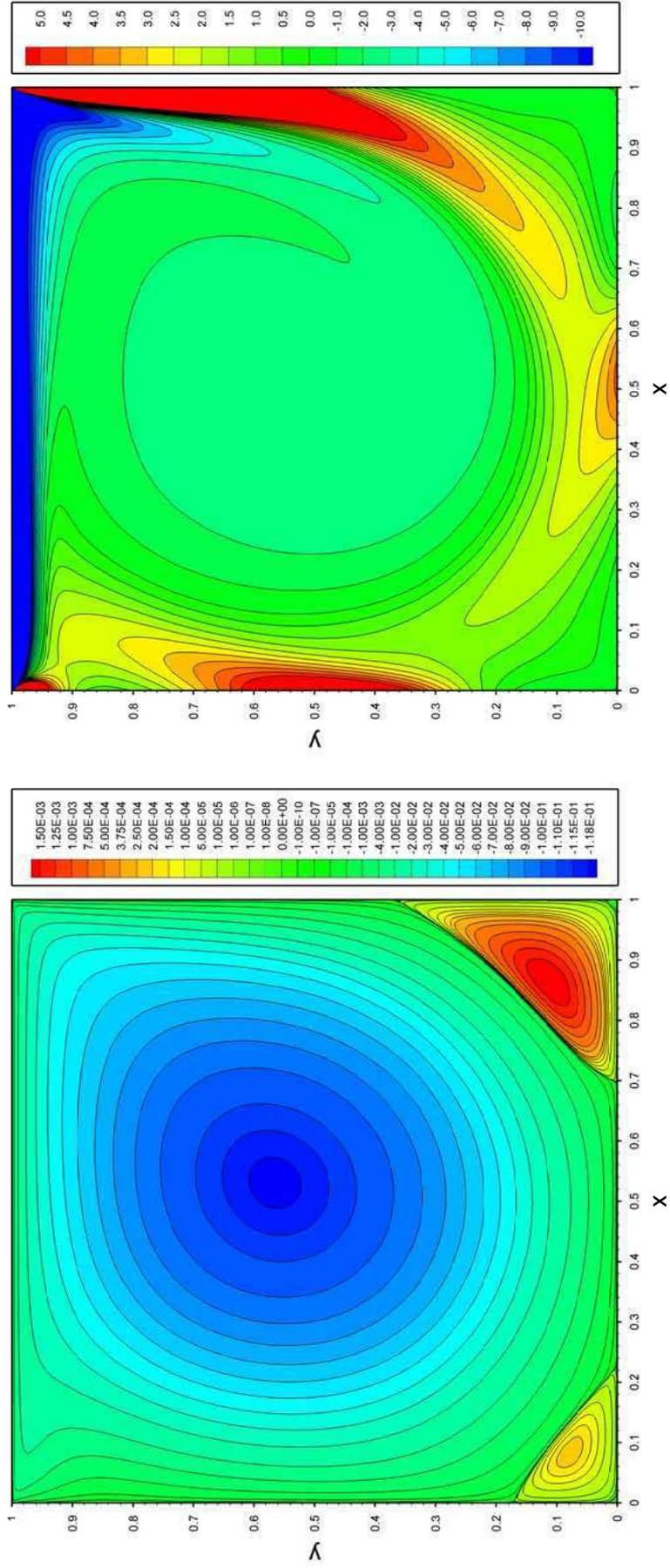

Figure 3) Streamline and vorticity contours of driven cavity flow, *Re*=1000.

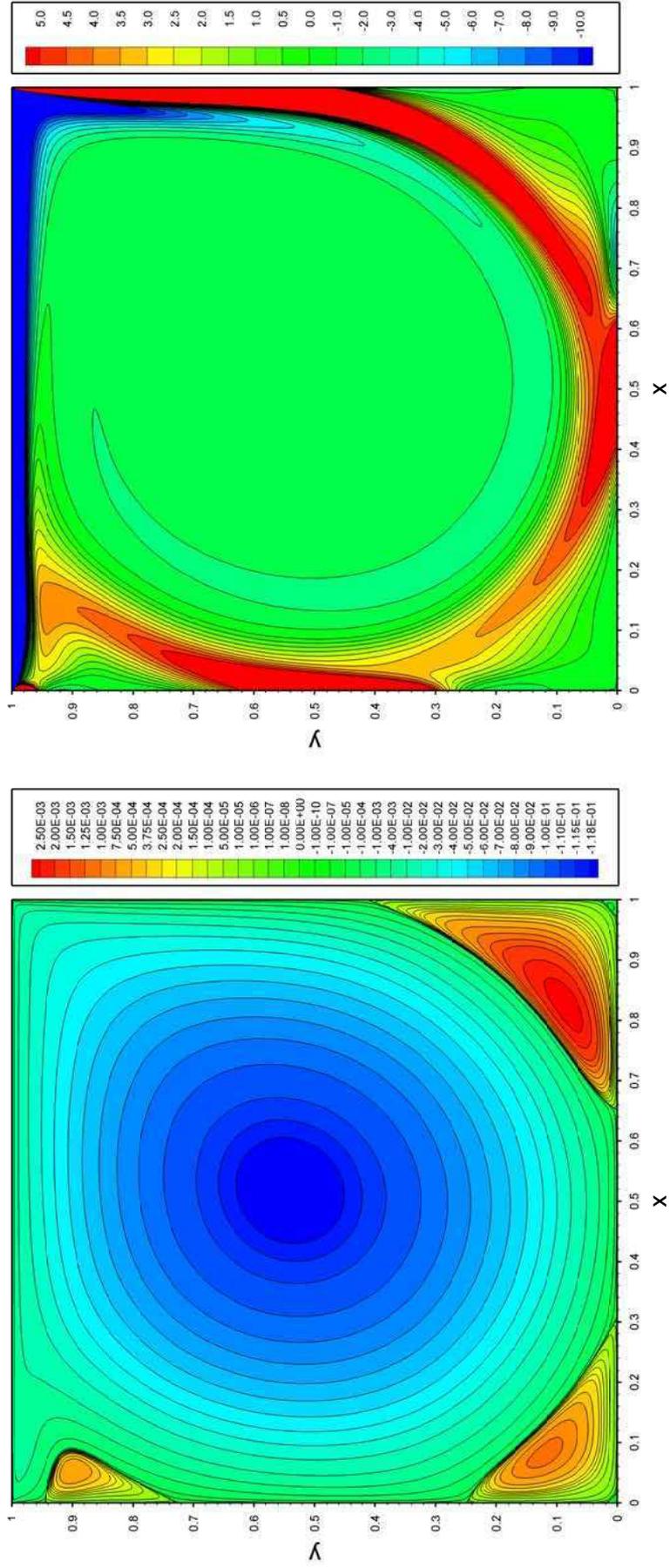

Figure 4) Streamline and vorticity contours of driven cavity flow, *Re*=3200.

| $\phi_x$ | $\dfrac{\phi_{i+1,j} - \phi_{i-1,j}}{2\Delta x}$ |
|---|---|
| $\phi_y$ | $\dfrac{\phi_{i,j+1} - \phi_{i,j-1}}{2\Delta y}$ |
| $\phi_{xx}$ | $\dfrac{\phi_{i+1,j} - 2\phi_{i,j} + \phi_{i-1,j}}{\Delta x^2}$ |
| $\phi_{yy}$ | $\dfrac{\phi_{i,j+1} - 2\phi_{i,j} + \phi_{i,j-1}}{\Delta y^2}$ |
| $\phi_{xy}$ | $\dfrac{\phi_{i+1,j+1} - \phi_{i-1,j+1} - \phi_{i+1,j-1} + \phi_{i-1,j-1}}{4\Delta x \Delta y}$ |
| $\phi_{xxy}$ | $\dfrac{\phi_{i+1,j+1} - 2\phi_{i,j+1} + \phi_{i-1,j+1} - \phi_{i+1,j-1} + 2\phi_{i,j-1} - \phi_{i-1,j-1}}{2\Delta x^2 \Delta y}$ |
| $\phi_{xyy}$ | $\dfrac{\phi_{i+1,j+1} - 2\phi_{i+1,j} + \phi_{i+1,j-1} - \phi_{i-1,j+1} + 2\phi_{i-1,j} - \phi_{i-1,j-1}}{2\Delta x \Delta y^2}$ |
| $\phi_{xxyy}$ | $\dfrac{\phi_{i+1,j+1} - 2\phi_{i,j+1} + \phi_{i-1,j+1} - 2\phi_{i+1,j} + 4\phi_{i,j} - 2\phi_{i-1,j} + \phi_{i+1,j-1} - 2\phi_{i,j-1} + \phi_{i-1,j-1}}{\Delta x^2 \Delta y^2}$ |

Table 1)  Standard second order central discretizations, $\mathcal{O}(\Delta x^2, \Delta y^2)$.

| Re | $\alpha$ | Second Order Accuracy | | Fourth Order Accuracy | | $\dfrac{\text{CPU}_{(\mathcal{O}\Delta x^2)}}{\text{CPU}_{(\mathcal{O}\Delta x^4)}}$ | $\dfrac{\text{Iteration No}_{(\mathcal{O}\Delta x^4)}}{\text{Iteration No}_{(\mathcal{O}\Delta x^2)}}$ |
| --- | --- | --- | --- | --- | --- | --- | --- |
| | | $\text{CPU}_{(\mathcal{O}\Delta x^2)}$ | $\text{Iteration No}_{(\mathcal{O}\Delta x^2)}$ | $\text{CPU}_{(\mathcal{O}\Delta x^4)}$ | $\text{Iteration No}_{(\mathcal{O}\Delta x^4)}$ | | |
| 100 | 0.6 | 182.57 | 25393 | 321.83 | 25091 | 1.76 | 0.99 |
| | 0.4 | 271.36 | 37739 | 478.90 | 37362 | 1.76 | 0.99 |
| | 0.2 | 531.62 | 73932 | 942.35 | 73506 | 1.77 | 0.99 |
| | 0.79 | 184.60 | 25677 | - | - | - | - |
| | 0.78 | 187.07 | 26013 | 336.72 | 26260 | 1.80 | 1.01 |
| | 0.7 | 207.94 | 28949 | 374.82 | 29215 | 1.80 | 1.01 |
| 1000 | 0.6 | 242.14 | 33682 | 435.98 | 34007 | 1.80 | 1.01 |
| | 0.5 | 289.38 | 40278 | 521.84 | 40688 | 1.80 | 1.01 |
| | 0.4 | 359.96 | 50116 | 649.25 | 50662 | 1.80 | 1.01 |
| | 0.2 | 707.90 | 98518 | 1279.43 | 99831 | 1.81 | 1.01 |
| | 0.1 | 1384.97 | 192629 | 2509.86 | 195820 | 1.81 | 1.02 |
| 3200 | 0.1 | 1261.09 | 175558 | 2398.06 | 187023 | 1.90 | 1.07 |

Table 2) Comparison of the CPU time and iteration number for second and fourth order accuracy obtained using ADI method

| $Re$ | $\alpha$ | Second Order Accuracy | | Fourth Order Accuracy | | $\dfrac{CPU_{(\mathcal{O}\Delta x^4)}}{CPU_{(\mathcal{O}\Delta x^2)}}$ | $\dfrac{\text{Iteration No}_{(\mathcal{O}\Delta x^4)}}{\text{Iteration No}_{(\mathcal{O}\Delta x^2)}}$ |
|---|---|---|---|---|---|---|---|
| | | $CPU_{(\mathcal{O}\Delta x^2)}$ | Iteration No$_{(\mathcal{O}\Delta x^2)}$ | $CPU_{(\mathcal{O}\Delta x^4)}$ | Iteration No$_{(\mathcal{O}\Delta x^4)}$ | | |
| 100 | 0.6 | 228.76 | 25419 | 367.15 | 25115 | 1.60 | 0.99 |
| | 0.4 | 339.95 | 37763 | 545.53 | 37386 | 1.60 | 0.99 |
| | 0.2 | 666.38 | 73953 | 1072.67 | 73528 | 1.61 | 0.99 |
| | 1.89 | 98.85 | 10963 | - | - | - | - |
| | 1.75 | 106.45 | 11820 | 176.40 | 12082 | 1.66 | 1.02 |
| | 1.6 | 115.97 | 12888 | 192.65 | 13191 | 1.66 | 1.02 |
| 1000 | 1.2 | 153.90 | 17091 | 254.16 | 17406 | 1.65 | 1.02 |
| | 0.6 | 303.67 | 33734 | 497.10 | 34067 | 1.64 | 1.01 |
| | 0.4 | 451.49 | 50165 | 741.81 | 50717 | 1.64 | 1.01 |
| | 0.2 | 888.35 | 98564 | 1457.44 | 99884 | 1.64 | 1.01 |
| | 0.1 | 1733.51 | 192673 | 2854.97 | 195871 | 1.65 | 1.02 |
| 3200 | 1.2 | 155.51 | 17290 | 261.89 | 17986 | 1.68 | 1.04 |

Table 3) Comparison of the CPU time and iteration number for second and fourth order accuracy obtained using Erturk method,

| $Re$ | $\alpha$ | Second Order Accuracy | | Fourth Order Accuracy | | $\dfrac{\text{CPU}_{(\mathcal{O}\Delta x^4)}}{\text{CPU}_{(\mathcal{O}\Delta x^2)}}$ | $\dfrac{\text{Iteration No}_{(\mathcal{O}\Delta x^4)}}{\text{Iteration No}_{(\mathcal{O}\Delta x^2)}}$ |
|---|---|---|---|---|---|---|---|
| | | $\text{CPU}_{(\mathcal{O}\Delta x^2)}$ | $\text{Iteration No}_{(\mathcal{O}\Delta x^2)}$ | $\text{CPU}_{(\mathcal{O}\Delta x^4)}$ | $\text{Iteration No}_{(\mathcal{O}\Delta x^4)}$ | | |
| 100 | 0.6 | 128.56 | 25393 | 165.61 | 25091 | 1.29 | 0.99 |
| | 0.4 | 191.07 | 37739 | 246.44 | 37362 | 1.29 | 0.99 |
| | 0.2 | 374.29 | 73932 | 484.75 | 73506 | 1.30 | 0.99 |
| 1000 | 0.79 | 130.40 | 25677 | - | - | - | - |
| | 0.78 | 131.81 | 26013 | 173.11 | 26260 | 1.31 | 1.01 |
| | 0.7 | 146.70 | 28949 | 192.82 | 29215 | 1.31 | 1.01 |
| | 0.6 | 170.49 | 33682 | 224.27 | 34007 | 1.32 | 1.01 |
| | 0.5 | 204.44 | 40278 | 267.97 | 40688 | 1.31 | 1.01 |
| | 0.4 | 253.79 | 50116 | 333.91 | 50662 | 1.32 | 1.01 |
| | 0.2 | 499.28 | 98518 | 658.93 | 99831 | 1.32 | 1.01 |
| | 0.1 | 975.98 | 192629 | 1291.06 | 195820 | 1.32 | 1.02 |
| 3200 | 0.1 | 889.31 | 175558 | 1233.30 | 187023 | 1.39 | 1.07 |

Table 4) Comparison of the CPU time and iteration number for second and fourth order accuracy obtained using ADI method, compiled with (-fast -O5) optimization

| $Re$ | $\alpha$ | Second Order Accuracy | | Fourth Order Accuracy | | $\dfrac{\text{CPU}_{(\mathcal{O}\Delta x^4)}}{\text{CPU}_{(\mathcal{O}\Delta x^2)}}$ | $\dfrac{\text{Iteration No}_{(\mathcal{O}\Delta x^4)}}{\text{Iteration No}_{(\mathcal{O}\Delta x^2)}}$ |
|---|---|---|---|---|---|---|---|
| | | $\text{CPU}_{(\mathcal{O}\Delta x^2)}$ | $\text{Iteration No}_{(\mathcal{O}\Delta x^2)}$ | $\text{CPU}_{(\mathcal{O}\Delta x^4)}$ | $\text{Iteration No}_{(\mathcal{O}\Delta x^4)}$ | | |
| 100 | 0.6 | 124.71 | 25419 | 165.94 | 25115 | 1.33 | 0.99 |
| | 0.4 | 184.99 | 37763 | 247.33 | 37386 | 1.34 | 0.99 |
| | 0.2 | 361.82 | 73953 | 486.49 | 73528 | 1.34 | 0.99 |
| 1000 | 1.89 | 53.51 | 10963 | - | - | - | - |
| | 1.75 | 58.01 | 11820 | 80.14 | 12082 | 1.38 | 1.02 |
| | 1.6 | 62.94 | 12888 | 87.18 | 13191 | 1.39 | 1.02 |
| | 1.2 | 83.98 | 17091 | 114.99 | 17406 | 1.37 | 1.02 |
| | 0.6 | 165.19 | 33734 | 226.39 | 34067 | 1.37 | 1.01 |
| | 0.4 | 245.57 | 50165 | 335.01 | 50717 | 1.36 | 1.01 |
| | 0.2 | 483.04 | 98564 | 661.49 | 99884 | 1.37 | 1.01 |
| | 0.1 | 937.74 | 192673 | 1284.31 | 195871 | 1.37 | 1.02 |
| 3200 | 1.2 | 83.91 | 17290 | 117.73 | 17986 | 1.40 | 1.04 |

Table 5) Comparison of the CPU time and iteration number for second and fourth order accuracy obtained using Erturk method, compiled with (`-fast -O5`) optimization